\documentclass[12pt,journal]{article}
\stepcounter{section}
\usepackage{graphicx,amssymb,amsmath,amsthm,cite,cases}
\usepackage{setspace}
\usepackage{color,multirow}
\usepackage{mathrsfs}
\def\oa{\hat{\operatorname{J}}}
\DeclareMathOperator*{\argmax}{arg\,max}
\def\be{\begin{equation}}
\def\ee{\end{equation}}
\def\ben{\begin{eqnarray}}
\def\een{\end{eqnarray}}

\def\D{\mathcal{D}}
\def\R{\mathbb{R}}

\def\F{\mathrm{F}}

\def\bsr{\mathrm{bsr}}
\def\bpp{\mathrm{bpp}}
\def\bweb{\mathrm{bwb}}
\def\bjp{\mathrm{bjp}}
\def\bjpp{\mathrm{bjp2}}

\def\til{\tilde}

\def\ctq{\til{c}^\Delta}

\def\vI{\mathbf{I}}
\def\vG{\mathbf{G}}

\def\vd{\mathbf{d}}

\def\vg{\mathbf{g}}

\def\vA{\mathbf{A}}
\def\vB{\mathbf{B}}
\def\vT{\mathbf{T}}
\def\vs{\mathbf{s}}

\def\st{\til{s}}
\def\on{o}
\def\ont{\til{\on}}

\def\vI{\mathbf{I}}
\def\vQ{\mathbf{Q}}
\def\vW{\mathbf{W}}

\def\vU{\mathbf{U}}

\def\vUkm{\vU^k\{m\}}
\def\vWkm{\vW^k\{m\}}
\def\vRe{\mathbf{R}}

\def\Ut{\til{U}}
\def\It{\til{I}}
\def\vIt{\til{\vI}}
\def\Wt{\til{W}}
\def\Wtkm{\til{W}^k\{m\}}
\def\vWtkm{\til{\vW}^k\{m\}}
\def\vUtkm{\til{\vU}^k\{m\}}
\def\Utkm{\til{U}^k\{m\}}

\def\vIm{{\vI}\{m\}}

\def\vd{\mathbf{d}}
\def\dwt{\mathbf{dwt}}
\def\idwt{\mathbf{idwt}}

\def\kq{k_q}
\def\Nq{N_b}
\def\ov{\overline}
\def\PSNR{\mathrm{PSNR}}
\def\mPSNR{\overline{\mathrm{PSNR}}}

\def\bpp{\mathrm{bpp}}

\def\MSE{\mathrm{MSE}}

\def\Mx{M_x}
\def\My{M_y}

\newcommand{\la}{\left \langle}
\newcommand{\ra}{\right \rangle}

\newtheorem{remark}{Remark}

\def\Nx{N_x}
\def\Ny{N_y}
\def\Nxy{\Nx \times \Ny}

\def\Mx{M_x}
\def\My{M_y}

\title{Enhancing sparse representation of color images  
 by cross channel transformation}
\author{Laura Rebollo-Neira\\ 
Mathematics Department\\
Aston University\\
B4 7ET, Birmingham, UK\\
Aurelien Inacio\\
ENSIIE, 
1 Rue de la R\'{e}sistance, \\
91000 \'{E}vry-Courcouronnes, France}
\begin{document}
\maketitle
\date{}
\begin{abstract}
Transformations for enhancing sparsity in 
the approximation of color images by 2D atomic 
decomposition are discussed. The sparsity is firstly 
considered with respect to the most significant coefficients 
in the wavelet decomposition of the color image. 
The discrete cosine transform is singled out as an effective 
transformation for this purpose. 
The enhanced feature is 
further exploited by approximating the transformed 
arrays using an effective greedy strategy with a
separable highly redundant dictionary. The relevance of 
the achieved sparsity is illustrated by a simple 
encoding procedure. On a set of typical test images 
the compression at high quality recovery is 
shown to significantly improve upon 
JPEG and WebP formats. The results are competitive with 
 those produced by the JPEG2000 standard. 
\end{abstract}
\maketitle
\section{Introduction}
In the signal processing field sparse representation usually refers to the approximation of a signal as superposition of elementary components, called atoms, which are members of a large redundant set, called a dictionary \cite{MZ93}. 
The superposition, termed atomic decomposition, 
 aims at approximating the signal involving as few atoms 
as possible 
\cite{MZ93,Mall09, Don01,EKB10}. 
Sparsity is also relevant to data collection. Within the 
 emerging theory of sampling known as {\em{compressive sensing}} (CS) 
 this property is key for reconstructing signals from a
 reduced number of measures \cite{CW08,Rom08, Bar11}. 
In particular, distributed compressive sensing (DCS) 
algorithms exploit inter signal correlation structure for 
multiple signal recovery \cite{DSB05,TZS21}  

Sparse image representation using redundant dictionaries
has been considered in numerous
 works e.g. 
\cite{WMM10, Ela10, ZXY15} and in the
 context of applications such as image restoration 
\cite{MES08,ZSL13,SZLi17}, feature extraction 
\cite{WYG09,YLY12,HVP19, RRB20} and
super resolution  \cite{YWH10,ZLY15, ZLL18, LCZ20}. 

The sparsity property of some types of 3D images
benefits from 3D processing \cite{CC13,CML15,DYK17,
MM17}. In particular most RGB (Red-Green-Blue)
images admit a sparser atomic decomposition
if approximated by 3D atoms \cite{RNW19}. Within the 3D
framework the gain in sparsity comes at expenses
of computational cost though. 

The purpose of this work is to show 
 that the application of 
  a transformation across the direction of the colors 
 improve sparsity in the wavelet domain representation of 
the 2D color channels. 
The relevance of this feature is demonstrated by 
a simple encoding procedure rendering good 
compression results in comparison to the state 
of the art. 
% comparison with the compression standards
%JPEG,  and JPEG2000.

The contributions of the paper are listed below.
\begin{itemize}

\item Highlighting of a distinctive feature 
concerning sparse representation of RGB images. Namely, 
that transformations in the direction of 
the color channels significantly improve the sparsity of 
the 2D atomic decomposition of the image.

\item Demonstration 
 that the discrete cosine transform (dct) 
 can improve sparsity with respect to the optimal 
 decorrelation transform, i.e. that given by the eigenvectors of the correlation matrix, which is thereby image dependent.

\item  Showcasing of the relevance of sparsity  to 
image compression with high quality 
recovery.  
On a set of 15 typical test images the achieved 
compression is shown to significantly improve upon JPEG and 
WebP formats. The results are competitive with those 
produced by JPEG2000. 
\end{itemize}

The work is organised as follows:
Sec.~\ref{notation}  introduces the mathematical 
notation. Sec.~\ref{decorre} compares several 
cross color transformations enhancing 
sparsity. 
A numerical example on a large data set is used to 
illustrate the suitability of the dct for that purpose. 
Sec.~\ref{atomdec} demonstrates the gain 
in sparsity obtained by the atomic decomposition of 
color images when the dct is applied  across the 
 RGB channels.
Sec.~\ref{ImComp} illustrates the relevance of the approach 
to image compression with high quality recovery. The conclusions are summarised in 
Sec.~\ref{Conclu}.

\section{Mathematical Notation}
\label{notation}
Throughout the paper we use the following 
notational convention.
$\R$ represents the set of real numbers.
Boldface letters indicate Euclidean vectors,
2D and 3D arrays.
Standard mathematical fonts indicate components,
e.g., $\vd \in \R^N$ is a vector of components
$d(i)\in \R,\, i=1,\ldots,N$. The  elements of a
 2D array $\vI \in \R^{\Nxy}$ are indicated as
$I(i,j),\,i=1,\ldots, \Nx, \, j=1,\ldots,\Ny$ and 
the color channels of $\vI \in \R^{\Nxy\times 3}$ as 
$I(:,:,z),\, z=1,2,3$. 
The transpose of a matrix, $\vG$ say, is
indicated as $\vG^\top$.
 A set of,  say $M$, color 
images is represented by the arrays  $\vI\{m\}\in \R^{\Nxy\times 3}, m=1,\ldots,M.$

The inner product between arrays in $\R^{\Nxy}$
 is given by the Frobenius inner product 
$\la \cdot, \cdot \ra_{\F} $ as 
$$\la \vG, \vI \ra_{\F}= \sum_{i=1}^{\Nx} \sum_{j=1}^{\Ny} 
 G(i,j)  I(i,j).$$ 
Consequently, the Frobenius norm $\| \cdot\|_{\F}$ is 
calculated as 
$$\|\vG\|_{\F}= \sqrt{\sum_{i=1}^{\Nx} \sum_{j=1}^{\Ny} 
 G(i,j)^2}.$$
The  inner produce between arrays in $\R^N$ is given by the 
Euclidean inner product  $\la \cdot, \cdot \ra$ 
 as
$$\la \vd ,\vg \ra= \sum_{i=1}^{N}
  d(i)g(i).$$

\section{Cross color transformations}
\label{decorre}
Given a color image 
$I(i,j,k), i=1,\dots,\Nx, j=1,\dots,\Ny, \, k=1,2,3$
the processing  of the 3 channels 
can be realised either in the  pixel/intensity 
or in the wavelet domain. Since 
the representation of most images is sparser in the 
wavelet domain \cite{GS01,RNMB13,LRN17,LRN18,RNW19} 
we approximate 
in that domain and reconstruct the approximated image 
by the inverse wavelet transform.
Thus, using a $3 \times 3$ matrix $\vT$,   
 we construct the transformed arrays $\vU$ and $\vW$
as follows
\ben
U(:,:,z)&\!\!=&\!\!\sum_{l=1}^3 I(:,:,l) T(l,z)\quad z=1,2,3. 
\label{Ua}\\
W(:,:,z)&\!\!=\!\!&\dwt(U(:,:,z))\,\,\,\,\,\,\,\quad 
z=1,2,3, 
\label{Wa}
\een
where $\dwt$ indicates the 2D wavelet transform. 
For the transformation $\vT$ we consider the following cases
\begin{itemize}
\item [(i)]  The dct.
\item [(ii)] The reversible YCbCr color space transform.
\item [iii)] The principal components (PC) transform.
\item [iv)] A transformation learned from an independent 
set of images.
\end{itemize}
The dct is given by the matrix
\begin{equation*}
\begin{pmatrix}
\frac{1}{\sqrt{3}} & \frac{\sqrt{2}}{\sqrt{3}}\cos(\frac{\pi}{6}) &  \frac{\sqrt{2}}{\sqrt{3}}\cos(\frac{\pi}{3}) \\
\frac{1}{\sqrt{3}} & 0 & \frac{\sqrt{2}}{\sqrt{3}}\cos(\pi) \\
\frac{1}{\sqrt{3}} & \frac{\sqrt{2}}{\sqrt{3}}\cos(\frac{5\pi}{6})&  \frac{\sqrt{2}}{\sqrt{3}}\cos(\frac{5 \pi}{6})
\end{pmatrix}
\end{equation*}
The YCbCr color space transform  
is given by the matrix below \cite{TM02} 
\begin{equation*}
\begin{pmatrix}
%0.299   & 0.587    &  0.114\\
%-0.16875& -0.33126 &  0.5\\
%0.5     & -0.41869 &-0.08131\\
0.299   & -0.169 & 0.5 \\
0.587   & -0.331 & -0.419\\
 0.114  & 0.5 & -0.0813\\
\end{pmatrix}
\end{equation*}     
The columns of the principal components transform are the 
normalised eigenvectors of covariance matrix of 
the RGB pixels. Thus, the transformation is image dependent and optimal with respect to decorrelating the color channels. 

As a first test, the approximation of the transformed
channels is realised by keeping a fixed number of the
largest absolute value entries, and setting the others
equal to zero.  In relation to this, for an image of 
size $L_x \times L_y \times 3$
we define the Sparsity Ratio (SR) as follows
\be
\label{SR}
\text{SR}=\frac{\text{$L_x \cdot L_y \cdot 3$}}{\text{Number of nonzero entries in the three channels}.}
\ee
The quality of a reconstructed image $\vIt$, with
respect to the original 8-bit image $\vI$, 
is compared using the Peak Signal-to-Noise Ratio ($\PSNR$),
\be
\label{psnr}
\PSNR=10 \log_{10}\left(\frac{255^2}{\MSE}\right),
\ee
where
$$
\MSE=\frac{1}{L_x \cdot L_y \cdot 3} 
\sum_{i,j,z=1}^{L_x,L_y,3}
(I(i,j,z)-\It(i,j,z))^2.
$$
For the numerical examples below the transformation 
corresponding to case (iv) is learned 
from a set of images $\vIm,\, m=1,\ldots,M$ all 
of the same size.% $L_x \times L_y \times 3$.  
  Starting from an invertible $3 \times 3$ 
 matrix $\vT^{k}$, with $k=1$, the learning algorithm 
proceeds through the following instructions.
\begin{itemize}
\item[1)]
Use $\vT^{k}$ to transform the arrays $\vIm \to \vUkm \to  
\vWkm$ as in \eqref{Ua} and \eqref{Wa}.
\item[2)]
Approximate each transformed array $\vWkm$ to obtain 
$\vWtkm$ by keeping the largest $K$ absolute value entries.
\item[3)]
Apply the inverse 2D wavelet transform $\idwt$ 
to reconstruct the approximated arrays  
$\vUtkm,\, m=1,\ldots,M$ as 
\ben
\label{UW2}
\Utkm(:,:,z)& \!\!\!=\!\!\!&\idwt(\Wtkm(:,:,z)),\, z=1,2,3. \nonumber
\een
\item[4)]
Use the original images $\vIm,\, m=1,\ldots,M$ to find  
%$\vG=(\vT^{k+1})^{-1}$ by least squares fitting, i.e 
$\vG=\vT^{-1}$ by least squares fitting, i.e 
$$\vG^\ast= \argmax_{\substack{G(z,l)\\z,l=1,2,3}} 
{\mathcal{E}}(\vG),$$
where
$${\mathcal{E}}(\vG)= \sum_{m=1}^M \sum_{\substack{i,j,l=1}}^{L_x, L_y,3} (I\{m\}(i,j,l)- \til{I}\{m\}(i,j,l))^2$$
with
$$ \til{I}\{m\}(i,j,l))= \sum_{z=1}^3 \Utkm(i,j,z) G(z,l).$$
\item [5)] While $\cal{E}$ decreases, or the 
maximum number of allowed iterations has not been reached, 
set 
$k \to k+1$, $\vT^{k}=(\vG^\ast)^{-1}$ and repeat steps 1) -- 5). 
\end{itemize}
Given the arrays $\vUtkm, \, m=1,\ldots,M$ the 
least squares problem for determining the transformation 
$\vT^k$ has a unique solution. 
However, the joint optimisation with respect to 
the arrays  $\vUtkm, m=1,\ldots,M$ {\em{and}} 
the transformation 
$\vT^k$ is not convex. Hence,  the 
 algorithm's outcome depends on the initial value.

\begin{figure}[h!]
\begin{center}
\hspace{-0.9cm}
\includegraphics[scale=0.5]{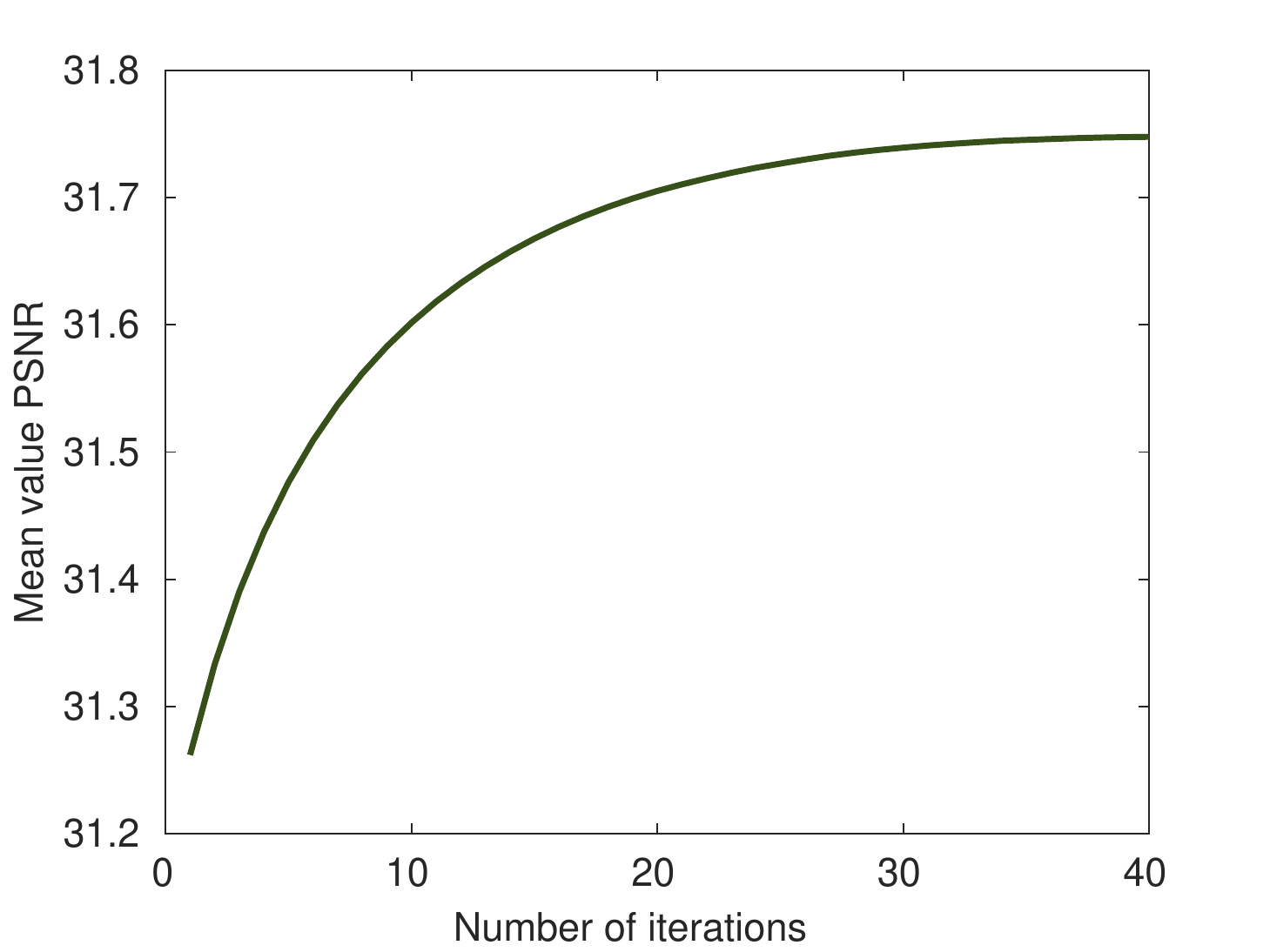} 
\includegraphics[scale=0.5]{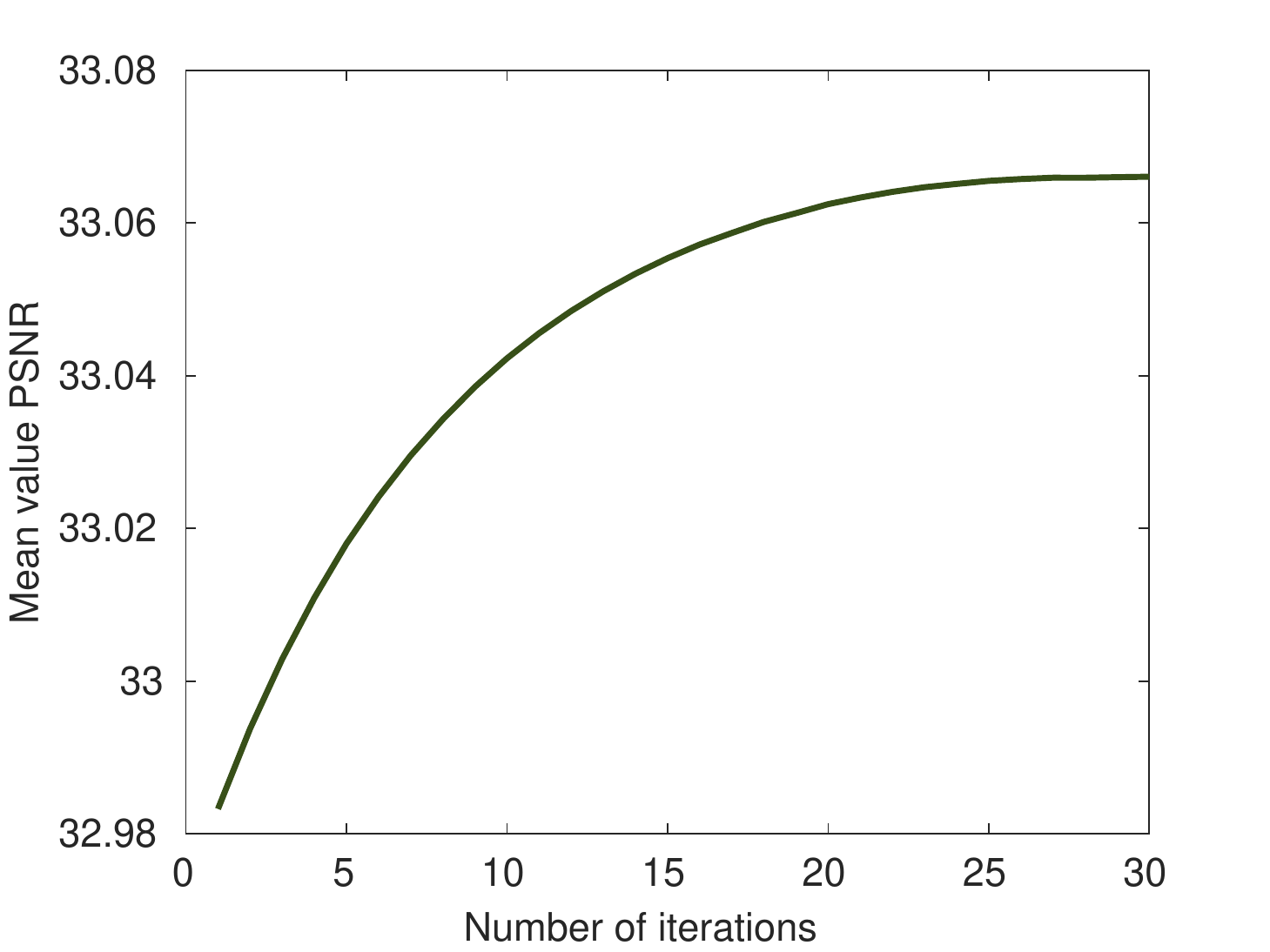}
\end{center}
\caption{Learning curves for the transformation 
(iv)  corresponding to two different random orthogonal 
transforms initialising the process. 
The mean value PSNR  
with respect to the 100 images in the training set 
corresponds to SR=15 for all the images.} 
\label{cur}
\end{figure}

 The transformation (iv)
  used in the numerical examples of Secs. 
\ref{NE1} and \ref{NE2} has been learned from $M=100$ 
images, 
all of size $384 \times 512 \times 3$,  from the    
UCID data set \cite{SS03}, which contains images of buildings, places and 
cars.
The learning curves for two random orthonormal 
initialisations is shown in Fig.~\ref{cur}.  

It is worth mentioning that, as shown in 
Fig. \ref{cur}, learning is richer when starting 
comparatively distant from a local 
 minimum (left graph in  Fig.~\ref{cur}). However, 
since the convergence is to a local minimum other 
random initialisations, even if generating  
 less learning, may give better results 
(right graph in  Fig.~\ref{cur}).

The aim of the next numerical test is to demonstrate 
the effect on the SR (c.f. \eqref{SR})  produced by  
 the above  (i) - (iv)  
transformations across the color channels.

%The approximation of the transformed
%channels is realised by keeping a fixed number of the
%largest absolute value entries, and setting the others
%equal to zero.  In relation to this
%we define the Sparsity Ratio (SR) as follows
%\be
%\label{SR}
%\text{SR}=\frac{\text{Number of elements in the arrays}}{\text{Number of nonzero entries}.}
%\ee
%
%The quality of a reconstructed image $\vIt$, with
%respect to the original 8-bit image $\vI$, 
%is compared using the Peak Signal-to-Noise Ratio ($\PSNR$),
%\be
%\label{psnr}
%\PSNR=10 \log_{10}\left(\frac{255^2}{\MSE}\right),
%\ee
%where
%$$
%\MSE=\frac{1}{3 L_x\cdot L_y} \sum_{i,j,z=1}^{L_x,L_y,3}
%(I(i,j,z)-\It(i,j,z))^2.
%$$
%
%improves on the sparsity of the image representation in 
%the wavelet domain. In other words, by keeping the 
%same number of entries in the transformed 
%2D arrays the quality of reconstructed  images is  
% higher if any of the three transformations being 
%considered is applied across the channels. 

\subsection{Numerical Test I}
\label{NE1}
\begin{figure}[h!]
\begin{center}
\hspace{-0.9cm}
\includegraphics[scale=0.8]{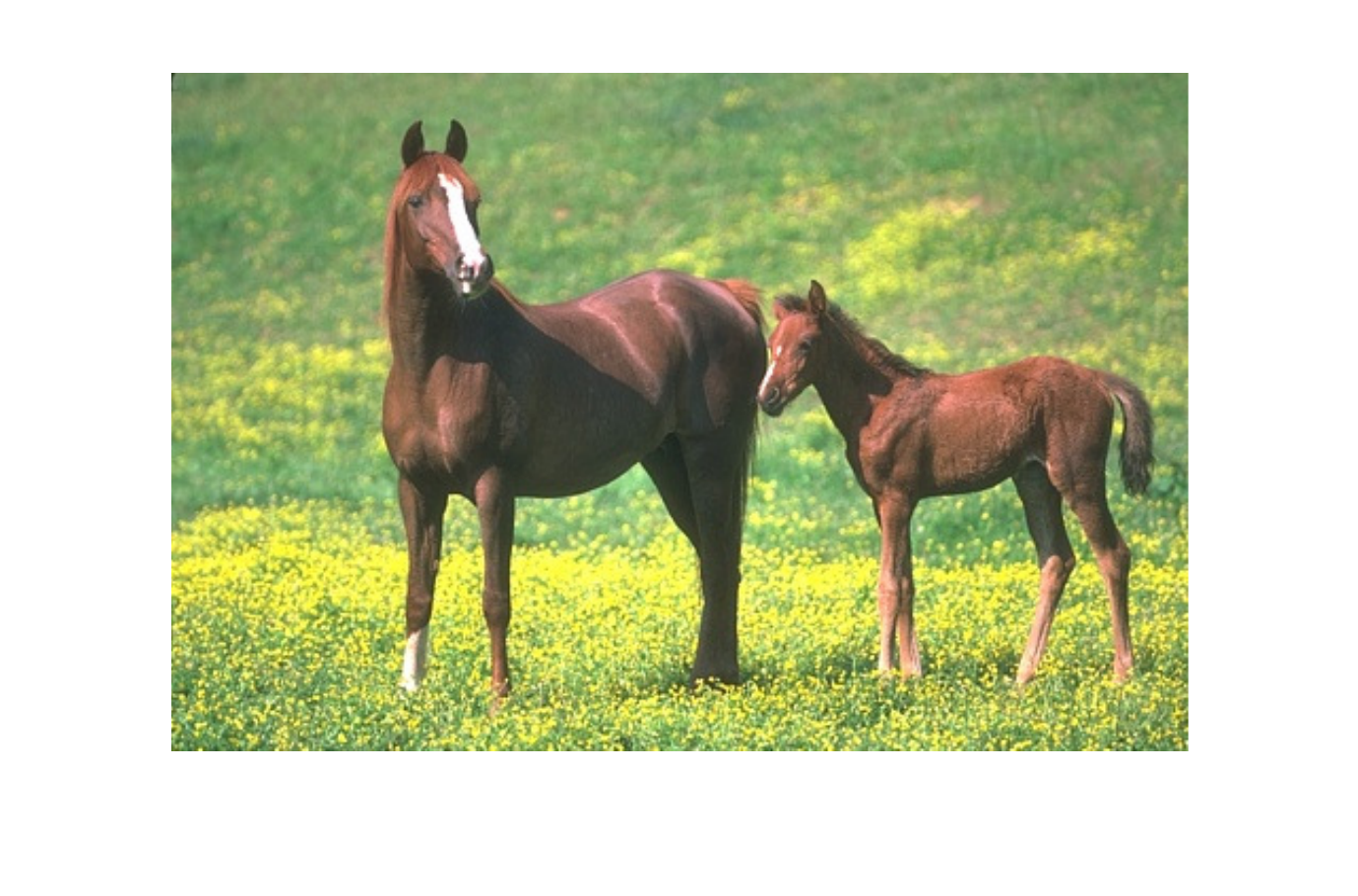}
\end{center}
\caption{One of the RGB images in the Berkeley's data set}
\label{OI}
\end{figure}

\begin{figure}[h!]
\begin{center}
\hspace{-1.3cm}
\includegraphics[scale=1.2]{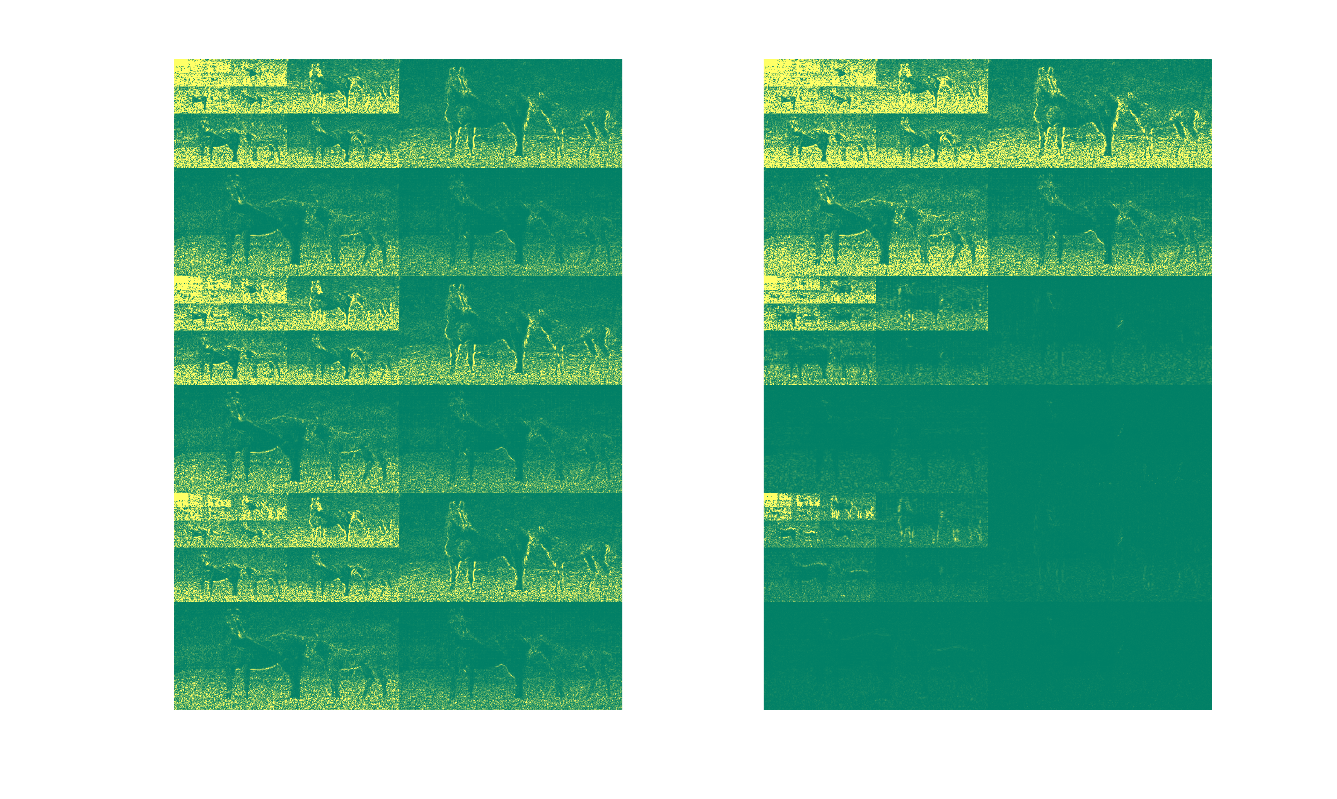} 
\end{center}
\caption{Magnitude of the entries in the array $\vW$ constructed as in \eqref{Wa} 
from the image of Fig.~\ref{OI}, with $\vT$ the dct (right 
graph) and without $\vT$ (left graph).}
\label{TW}
\end{figure}

Using the whole Berkeley data set \cite{Berk}, 
consisting of 300 images all of size  $321 \times 481 \times 3 $  we proceed with each image as in 
\eqref{Ua} and \eqref{Wa}. The $\dwt$ corresponds to 
 the Cohen-Daubechies-Feauveau 9$/$7 (CDF97) wavelet family.
 Fig.~\ref{TW}  shows the 
transformed channels of the image in Fig.\ref{OI}, 
including the dct transformation 
across channels (right graph) and without 
$\vT$ transformation (left graph). 

For the numerical test the approximations  are 
realised fixing SR=20 and SR=10.
The reconstructed images are obtained for the 
approximated arrays $\Wt$ as
\ben
\Ut(:,:,z)&=&\idwt(\Wt(:,:,z)),\quad \;\;\;\;\;\;\; z=1,2,3 
\label{iUa}\\
\It(:,:,z)&=&\!\!\!\sum_{k=1}^3\Ut(:,:,k) (T^{-1}(z,k)),\; z=1,2,3 
\label{iUb}
\een
where $\vT^{-1}$ is the inverse of $\vT$. When no 
transformation $\vT$ is considered the image is 
reconstructed directly from \eqref{iUa}.  

The 2nd and 4th columns of Table 
\ref{TABLE1} 
show the mean value $\PSNR$ ($\mPSNR$), with respect 
to the whole data set, for 
SR=20 and SR=10 respectively, corresponding to the 
transformations $\vT$ listed in the 1st column of the 
table.  
The 3rd and 5th columns give the standard deviations
(std). 

\begin{table}
%\label{TABLE1}
\begin{center}
\begin{tabular}{|l||r|r|r|r||}
\hline
Transf. & $\mPSNR$ & std& $\mPSNR$ & std \\ \hline
(i)\, dct  & 33.9 & 5.0  & 38.9 & 5.1  \\ \hline
(ii) YCbCr & 33.7 & 5.0  &38.7 & 5.1 \\ \hline
(iii) PC  & 33.7 & 4.9  &38.4 & 5.0  \\ \hline
(iv) Learned   & 33.7 & 4.9  &38.7 & 5.0 \\ \hline
(v) No transf. & 29.7 & 4.8 &32.9 & 5.2 \\ \hline \hline
\end{tabular}
\end{center}
\caption{Mean value PSNR, with respect to the 300
images in the
Berkeley data set. The  approximation of each image
is realised by setting the least significant entries
 in the arrays $\vW\{m\}=1,\ldots,300$ equal zero,
in order to obtain SR=20 for all the images  
(2nd column) and SR=10 for all the images
 (4th column). }
\label{TABLE1}
\end{table}

While Table \ref{TABLE1} shows that all (i) - (iv) 
transformations render equivalent superior results, 
with respect to not applying a transformation (case (v)), the dct slightly  exceeds 
 the others.
Case (iv) refers to the best results achieved
when initialising the learning algorithm with
 500 different random orthonormal transformations. 
When initialising the algorithm with transformations 
(i) - (iii) it appears that each of these transformations 
 is close to a local minimiser of the algorithm. 
This stems from the fact that 
 such initialisations do not generate significant learning.

The common feature of  most of the 300 images in the data set used in this numerical example is the correlation property of the three color channels. This property   
is  assessed by the correlation coefficients 
\ben
r(z)&=&\frac{\sum_{i=1}^{L_x}\sum_{j=1}^{L_y}\Gamma(i,j,z)\Gamma(i,j,z+1)}{\sigma(z) \sigma(z+1)},\quad z=1,2,\nonumber\\
r(3)&=&\frac{\Gamma(1)\Gamma(3)}{\sigma(1) \sigma(3)},\nonumber
\een
where
\ben
\Gamma(i,j,z)&\!\!=\!\!&(I(i,j,z)- \ov{I(:,:,z)}),\\ 
\sigma(z)&\!\!=&\!\!\sqrt{\sum_{i=1}^{L_x}\sum_{j=1}^{L_y} (I(i,j,z)- \ov{I(:,:,z)})^2},
\een
and $\ov{I(:,:,z)}$ indicates the mean value of 
channel $z$. 

\begin{figure}[h!]
\begin{center}
%\includegraphics[width=8cm]{OImage.eps}
%\hspace{-0.9cm}
\includegraphics[scale=0.5]{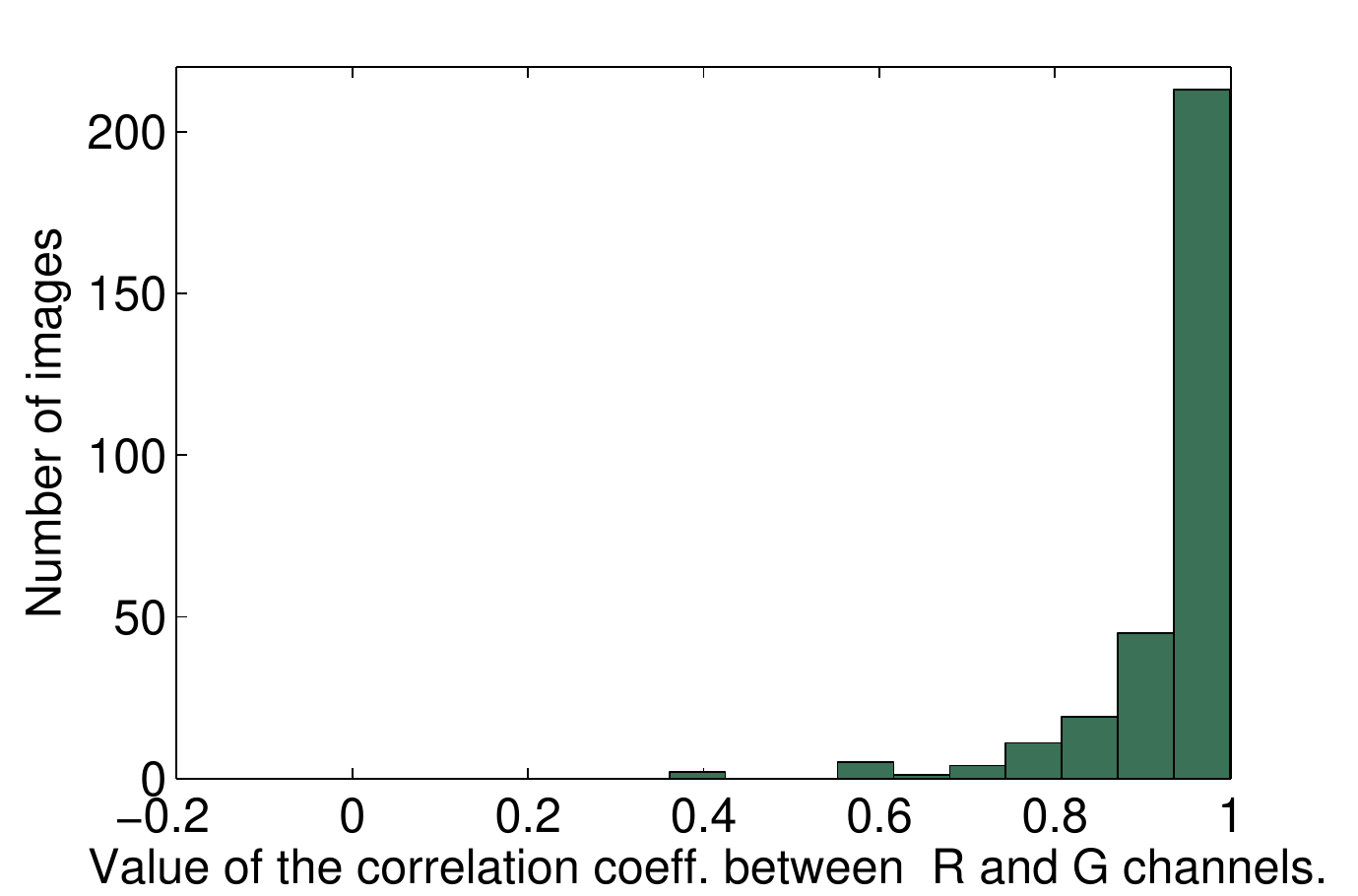}
\includegraphics[scale=0.5]{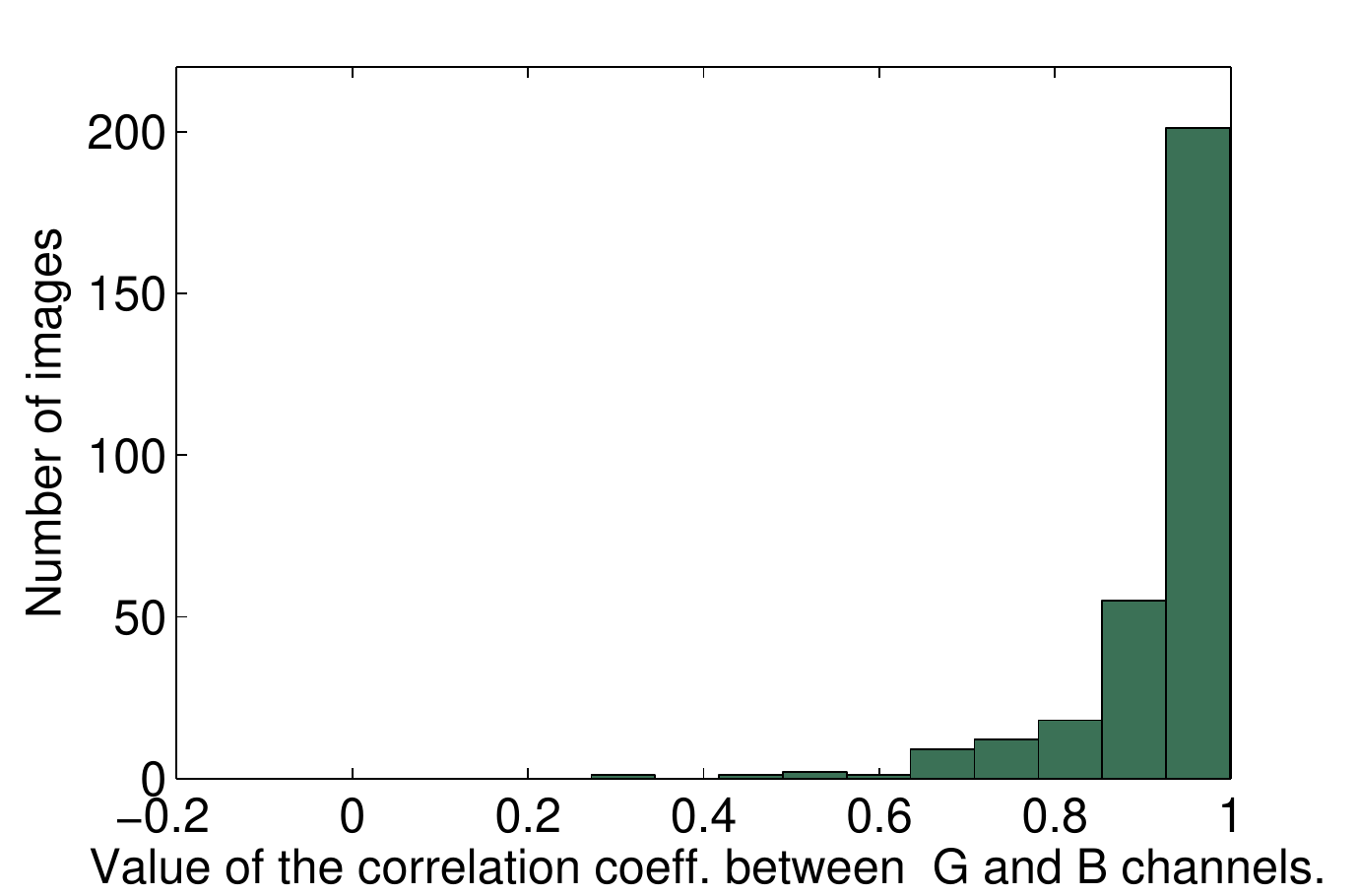}
\includegraphics[scale=0.5]{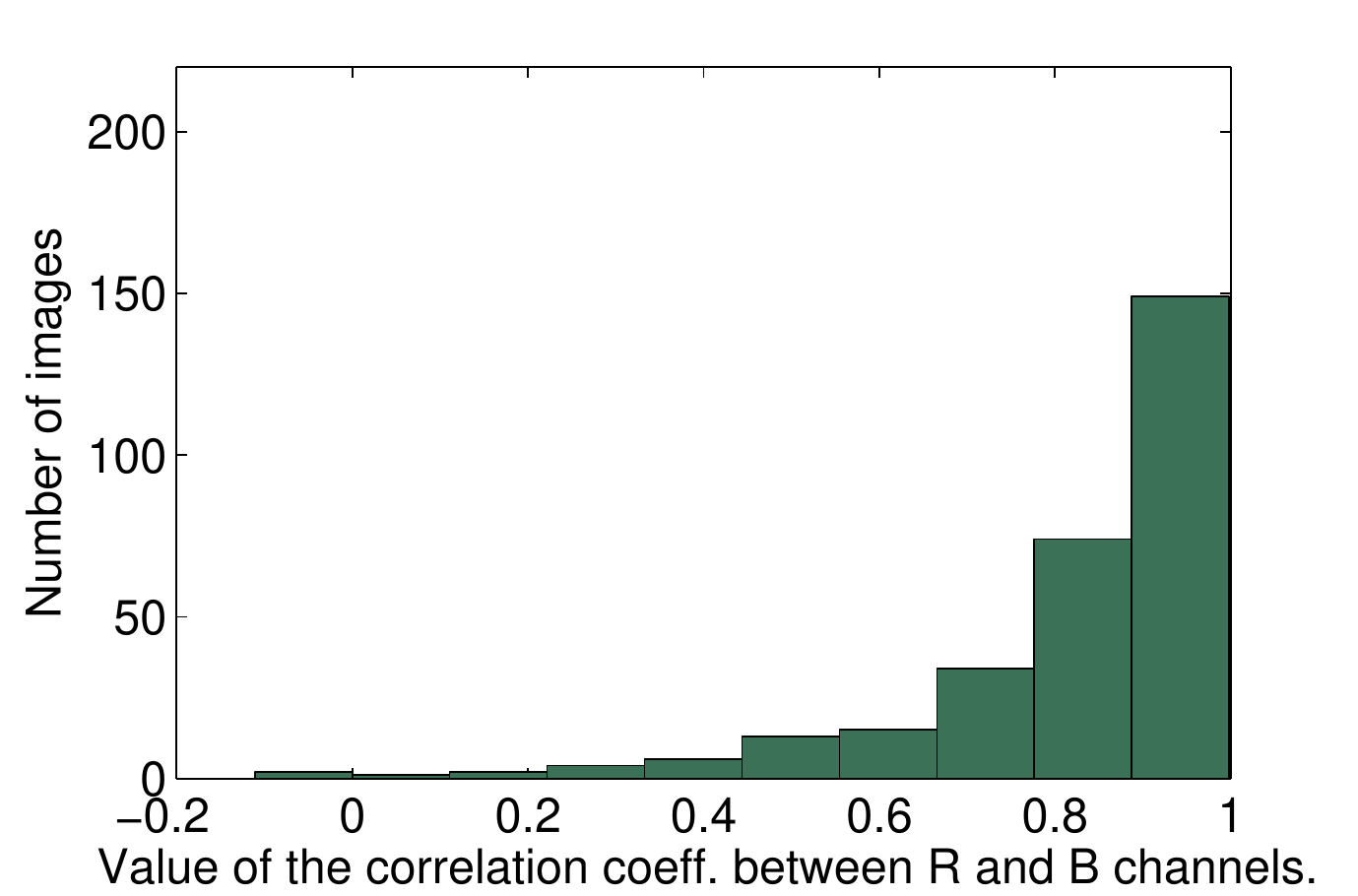}\\
\end{center}
\caption{Histograms of the correlation 
coefficients between the RGB channels.}
\label{corr3D}
\end{figure}

As seen in the 3 histograms of 
Fig.~\ref{corr3D} for most images in the data set 
the correlation of the color channels is high. It is 
surprising then than the PC transformation (iii), which 
completely decorrelates the channels, does not overperform 
  the other transformations, on the contrary.  
This feature has also been noticed 
in the context of bit allocation for subband color
 image compression \cite{GP07}. 

%In the next section we consider a dictionary 
%based approach for improving the quality of the
%approximation. 

\section{Approximations by atomic decomposition}
\label{atomdec}
We have seen (c.f. Table~\ref{TABLE1}) 
that by transformation of channels it is 
possible to gain quality when reducing nonzero 
entries in the 
channels. Now we discuss how to 
 improve quality further by approximating the
 2D arrays \eqref{Wa} by an atomic decomposition, 
 other than just by  
neglecting their less significant entries. 
For the approximation to be successful 
it is important to use an appropriate 
dictionary. To this end, one possibility  
could be to learn the dictionary from training data
 \cite{KMR03, AEB06, RZE10, TF11, ZGK11, HSK13, SNS15, WEB15}. However as demonstrated in previous works 
\cite{LRN17, LRN18, RNW19}
 a separable dictionary, which is easy to construct, 
is well suited for the purposes of 
achieving sparsity and delivers a fast implementation of 
the approach. 
 Since we use that dictionary in the numerical examples,  
below we describe the method 
for constructing the atomic decomposition of the array 
$\vW$ considering specifically a separable dictionary. 

Firstly we concatenate  the 3 planes $W(:,:,z), z=1,2,3$
into an extended 2D array $\vW'\in \R^{3 L_x \times L_y}$
 and divide this array 
 in small blocks $\vW'_q,\,q=1,\ldots,Q$. 
 Without loss of generality
the blocks are assumed to be square
of size $\Nq \times \Nq$ say, and are approximated 
using separable dictionaries 
$\D^x =\{\vd^x_n \in \R^{\Nq},\,\|\vd^x_n\|_2=1\}_{n=1}^{\Mx}$  and
$\D^y =\{\vd^y_m \in \R^{\Nq},\, \|\vd^y_n\|_2=1\}_{m=1}^{\My}$,
with $\Mx \My = M$.

For $q=1,\ldots,Q$ every element ${\vW'}_q$ is approximated
by an {\em{atomic decomposition}} as below:
\be
\label{atoq}
{\vW'}_q^{k_q}= \sum_{n=1}^{k_q}
c^{k_q,q}(n) \vd^x_{\ell^{x,q}_n}(\vd^y_{\ell^{y_,q}_n})^T,
\ee
where 
%$(\vd^y_{\ell^{y_,q}_n})^T$ indicates the
 ${\ell^{y_,q}_n}$ is the index 
in the set $\{1,2,\ldots,M_y\}$ corresponding to the
label of the atom in the dictionary $\D^y$ 
contributing to the
$n$-th term in the approximation of the $q$-th block.
The index ${\ell^{x_,q}_n}$ has the equivalent description. 
The assembling of the approximated blocks gives rise to 
the approximated array $\vW'^{K}$, i.e.,
${\vW'}^{K}= \oa_{q=1}^Q {\vW'}_q^{k_q}$,
where $K=\sum_{q=1}^Q k_q$ and $\oa$ represents  the
assembling operation. 

The approximation of the partition ${\vW'}_q=1,\ldots,Q$ 
 is carried out
iteratively as a two step process which selects  i) the 
 atoms in the atomic decomposition  
\eqref{atoq} and 
ii) the sequence in which the blocks in the partition
 are approximated.
The procedure is called Hierarchized Block Wise (HBW) 
implementation of
greedy pursuit strategies \cite{RNMB13, LRN16}. 
For the selection of the atoms we apply the 
Orthogonal Matching Pursuit (OMP) approach \cite{PRK93}
dedicated to 2D with separable dictionaries (OMP2D) 
\cite{RNBC12}. 
Thus, the whole algorithm is  termed 
HBW-OMP2D \cite{RNMB13}. 
The method iterates as described below.

On setting $\kq=0$ and  $\vRe_q^{0}={\vW'}_q$ at
iteration $\kq+1$ the algorithm selects the indices
$\ell^{x,q}_{\kq+1}$ and $\ell^{y,q}_{\kq+1}$,
as follows:
\be
\label{selec}
\ell^{x,q}_{\kq+1},\ell^{y,q}_{\kq+1}= \operatorname*{arg\,max}_{\substack{n=1,\ldots,\Mx\\
m=1,\ldots,\My}} \left |\la \vd^{x}_n ,\vRe_q^{\kq} \vd^{y}_m \ra_F \right|,\,\,\, q=1,\ldots,Q,
\ee
where $\vRe_q^{\kq}= {\vW'}_q - {\vW'}_q^{\kq}$.
The  calculation of ${\vW'}_q^{\kq}$ is realised 
in order to minimise $\|\vRe_q^{\kq}\|_F$, which is 
equivalent to finding the 
 orthogonal
projection onto the subspace spanned by the 
selected atoms$\{\vA_n=\vd^{x}_{\ell^{x,q}_n}  
(\vd^{y}_{\ell^{y,q}_n})^T\}_{n=1}^{\kq}$. 
 In our implementation, 
the calculation of the
coefficients $c^{q}(n),\,n=1,\ldots,\kq$ in \eqref{atoq} 
is realised as
\be
\label{Bcoe}
c^{k_q,q}(n)= \la \vB_n^k, {\vW'}_q^{k_q}\ra_\F,\quad n=1,\ldots,\kq,
\ee
where the set $\{\vB_n^{\kq}\}_{n=1}^{\kq}$ is 
biorthogonal to the set $\{\vA_n\}_{n=1}^{\kq}$ 
and needs to be  upgraded and updated to account for 
each newly selected atom. Starting from
$\vB_1^1= \vQ_1=\vA_{1}=\vd^x_{\ell^x_1} (\vd^y_{\ell^y_1})^\top$ the  updating and upgrading is realised through
 the  recursive equations \cite{RNBC12, RNL02}:
\be
\begin{split}
\vB_n^{\kq+1}&= \vB_n^{\kq} - \vB_{\kq+1}^{\kq+1}\la \vA_{\kq+1}, \vB_n^k \ra_\F,\quad n=1,\ldots,\kq\\
\vB_{\kq+1}^{\kq+1}&= \vQ_{\kq+1}/\|\vQ_{\kq+1}\|_\F^2,\,\,
 \text{where}\\
 \vQ_{\kq+1}&= \vA_{\kq+1} - \sum_{n=1}^{\kq} \frac{\vQ_n}
 {\|\vQ_n\|_\F^2} \la \vQ_n, \vA_{\kq+1}\ra_\F, 
 \end{split}
 \ee
with the additional re-orthogonalisation step 
\be
\label{GS}
\vQ_{\kq+1} \leftarrow  \vQ_{\kq+1}- \sum_{n=1}^{\kq} \frac{\vQ_n}{\|\vQ_n\|_\F^2}.
\la \vQ_n , \vQ_{\kq+1}\ra_\F.
\ee
As discussed in \cite{RNMB13, LRN16},  for 
$\ell^{x,q}_{\kq+1}$ and
$\ell^{y,q}_{\kq+1},\,q=1,\ldots,Q$  the
indices resulting from \eqref{selec}, the block
to be approximated in the next iteration
corresponds to the value $q^\star$ such that
$$q^\star= 
\operatorname*{arg\,max}_{q=1,\ldots,Q} 
\left |\la \vd^{x}_{\ell^{x,q}_{\kq+1}}, \vRe_q^{\kq} \,
\vd^{y}_{\ell^{y,q}_{\kq+1}} \ra\right|.
$$
The algorithm stops when the required total number of $K$ 
atoms has been selected. This number can be fixed using 
the SR, which is now calculated as
\be
\text{SR}=\frac{\text{Number of elements in the arrays}}{K}.
\ee
\begin{remark} 
The above described implementation of HBW-OMP2D is 
 very effective in terms of speed, but demanding 
in terms of memory (the partial outputs corresponding 
to all the blocks in the partition need to be stored 
at every iteration). 
An alternative implementation, termed 
HBW Self Projected Matching Pursuit (HBW-SPMP) 
\cite{LRN17,RNRS20}, would 
 enable the application of the identical strategy 
to much larger images than the ones considered in this work.
\end{remark}

\subsection{Numerical example II} 
\label{NE2}
\begin{figure}[h!]
\begin{center}
%\vspace{-1cm}
\includegraphics[width=9cm]{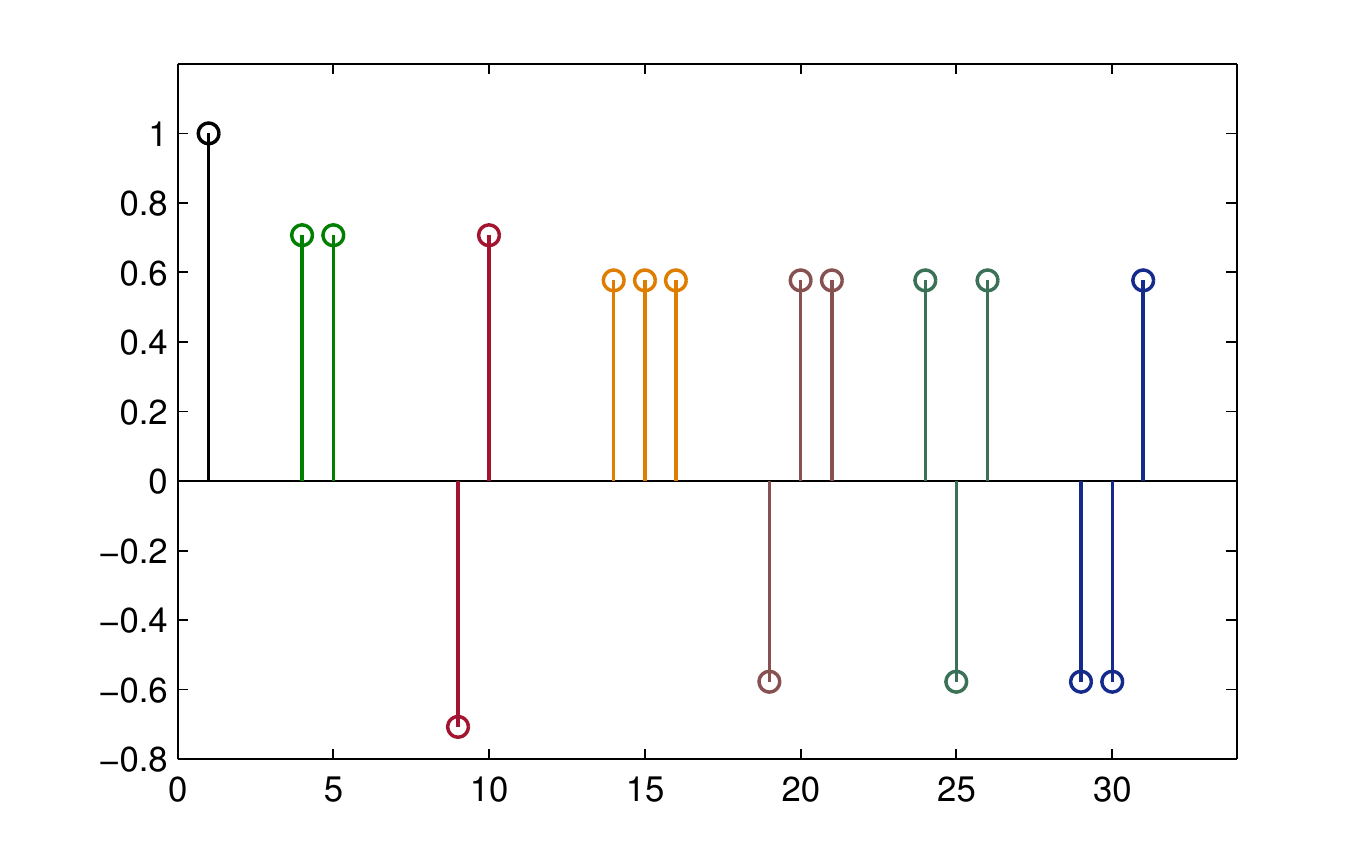}% \hspace{-0.1cm}
\end{center}
%\vspace{-1cm}
\caption{{\small{Prototypes (each in different color) that generate the
dictionaries $\mathcal{D}_{L}^x$ by sequential 
translations of one point.}}}
\label{protf}
\end{figure}

For this and the next numerical example, we 
use a mixed dictionary 
consisting of two classes of sub-dictionaries of
different nature:
\begin{itemize}
\item[I)]
 The trigonometric dictionaries  $\mathcal{D}_{C}^x$ and
$\mathcal{D}_{S}^x$, defined below,
\ben
\mathcal{D}_{C}^x&\!\!\!\!=\!\!\!\!\!\!&\{w_c(n)
\cos{\frac{{\pi(2i-1)(n-1)}}{2M}},i=1,\ldots,\Nq\}_{n=1}^{M}\nonumber\\
\mathcal{D}_{S}^x&\!\!\!\!=\!\!\!\!\!\!&\{w_s(n)\sin{\frac{{\pi(2i-1)(n)}}{2M}},i=1,\ldots,\Nq\}_{n=1}^{M},\nonumber
\een
where $w_c(n)$ and $w_s(n),\, n=1,\ldots,M$ are 
normalisation factors.
\item[II)]
The dictionary $\mathcal{D}_{L}^x$,
which is constructed
by translation of the prototype atoms in Fig.~\ref{protf}.
\end{itemize}
The mixed dictionary $\mathcal{D}^x$
is built as
$\mathcal{D}^x = \mathcal{D}_{C}^x \cup \mathcal{D}_{S}^x
\cup \mathcal{D}_{L}^x$ and
$\mathcal{D}^y= \mathcal{D}^x$. 

Table \ref{TABLE2} shows 
the improvement in $\mPSNR$ achieved 
by atomic decompositions using the mixed dictionary 
for SR= 20 and SR= 10. 
\begin{table}[h!]
%\label{TABLE2}
\begin{center}
\begin{tabular}{|l||r|r| r| r||}
\hline
 Transf. & $\mPSNR$ & std & $\mPSNR$ & std\\ \hline
(i)\, dct  & 40.5 & 5.0 & 48.1 & 4.4  \\ \hline
(ii) YCbCr  & 40.3 & 5.0  & 47.8 & 4.4\\ \hline
(iii) PC  & 39.6 & 4.8 & 46.3 & 4.5 \\ \hline
(iv) Learned  & 40.3 & 4.9& 47.4 & 4.3  \\ \hline
(v) No transf.  & 34.0 & 5.2& 39.1 & 5.6 \\ \hline \hline
\end{tabular}
\end{center}
\caption{Mean value PSNR, with respect to 300 images in the
Berkeley data set, produced  by 2D atomic decomposition
of the arrays $\vW\{m\}=1,\ldots,300$ in order to obtain
SR=20 (2nd column) and SR=10 (4th column).}
\label{TABLE2}
\end{table}

Notice that while 
case (v), which does not include any $\vT$ transformation,  
gives superior results than by disregarding 
entries (c.f. Table \ref{TABLE1}) 
 when applying any of the transformations
(i) -- (iv) the results improve further.  
The PC transform, however, appears
 significantly less effective than the others. 
In addition to rendering the best results, the dct brings 
along the additional advantage of being orthonormal. 
Consequently, it does not magnify errors at the 
inversion step. Because of this, we single out the dct 
as the most convenient cross color transformation out of 
the four considered here.

\section{Application to image compression}
\label{ImComp}

In order achieve compression by filing an 
atomic decomposition we need to address two 
issues. Namely, the quantisation of the coefficients
$c_{q}(n)\,n=1,\ldots,\kq,\, q=1,\ldots,Q$ in 
\eqref{atoq} and the storage of the indices 
$(\ell^{x,q}_n, \ell^{y,q}_n),n=1,\ldots,\kq,\, q=1,\ldots,Q$.  We tackle the matters by simple but effective procedures 
\cite{LRN18}. 

For $q=1,\ldots,Q$
the absolute value coefficients
$|c_q(n)|,\,  n=1,\ldots,\kq$ are converted to
integers through uniform quantisation as follows
\be
\label{uniq}
c^\Delta_{q}(n)=
\begin{cases}
\lceil \frac{|c_{q}(n)| - \theta }{\Delta} \rceil, &\mbox{if}% \quad \lceil  \frac{|c_{q}(n)| - \theta }{\Delta} \rceil \ge 0 \\ 
 \quad |c_{q}(n)| \ge \theta\\
 & 0 \quad \mbox{otherwise}.
\end{cases}
\ee
The signs of the coefficient are encoded separately as a
vector, $\vs_q$,
using a binary alphabet.
Each pair of indices $(\ell^{x,q}_n, \ell^{y,q}_n)$
corresponding to the atoms in the decompositions of the 
  block ${\vW}'_q$
 is mapped into a single index $\on_q(n)$.
The set $\on_q(1),\ldots,\on_q(\kq)$
 is sorted in ascending order
$\on_q(n)\rightarrow \ont_q(n),\,n=1,\ldots,\kq$
to take the differences $\delta_q(n)=\ont_q(n)- \ont_q(n-1),\,n=2,\ldots,\kq$ and construct the string of non-negative
 numbers ${\ont_q(1), \delta_q(2), \ldots, \delta_q(\kq)}$.
The order of the set $\ont_q(n),\,n=1,\ldots,\kq$
 induces
 order in the unsigned coefficients,
$c^\Delta_{q}(n) \rightarrow \ctq_{q}(n)$,
and in the  corresponding signs $s_{q}(n) \rightarrow
\st_{q}(n)$. 

For each $q$ the number 0 is added at the end
of the indices  $\ont_q(n),\,n=1,\ldots,\kq$
before concatenation, to be able to
 separate strings
corresponding to different blocks.
Each sequence of strings corresponding to $q=1,\ldots,Q$
is concatenated and encoded using
the off-the-shelf MATLAB function
{\tt{Huff06}} \cite{Karl}, which implements
Huffman coding.

 The compression rate is given in bits-per-pixel (bpp)
 which is defined as $$\bpp=\frac{\text{Size of the file in bits}}
{\text{Number of intensity pixels in a single
channel}}.$$

At the reconstruction stage the
indices $(\til{\ell}^{x,q}_n, \til{\ell}^{y,q}_n),\,
n=1\,\ldots \kq$  are  recovered  from the string of
differences $\delta_q(n)),\,n=2,\ldots,\kq$. 
 The signs of the coefficients are
read from the binary string.
The quantised unsigned
coefficients are read and transformed into real numbers as:
$$|\til{c}_{q}^{\mathrm{r}}(n)|= 
\Delta \cdot {\til{c}}_{q}^\Delta(n) + (\theta-\Delta/2)\,
\quad n=1\,\ldots \kq.$$

The codec for reproducing the examples in the 
next sections has been made available on 
\cite{webpage}.

\subsection{Numerical Example III}

The relevance to image compression of the achieved 
 sparsity by dct  cross color transformation  
 is illustrated in this section by comparison 
 with results yielded by the compression standards JPEG, 
WebP, and JPEG2000, on the 15 images in Table~\ref{TABLE3}. 
These are 
typical images, used for compression  tests,  
 available in ppm or png format. 
The first 9 images are classic test images 
taken from \cite{Iclassic}.  
The last six images are 
portions of $1024 \times 1024 \times 3$ pixels 
from very large high resolution  images available on
 \cite{ICom}. 

All the results have been obtained in the MATLAB environment
(version R2019a), using a machine with CPU 
%Intel Core I5 7300HQ and RAM 12Go DDR4.
Intel(R) Core(TM) i7-3520M RAM 8GB CPU @ 2.90GHz. 
 For the image approximation the 
HBW-OMP2D method was implemented with a 
 C++ MEX file. All the channels and all the images 
were partitioned into blocks of size $16 \times 16$. 
The approximation times to produce the 
results in Table~\ref{TABLE4} are displayed  in the last 
column of Table~\ref{TABLE3}.

\begin{figure}
\centering
%\vspace{-0.9cm}
\includegraphics[scale=0.14]{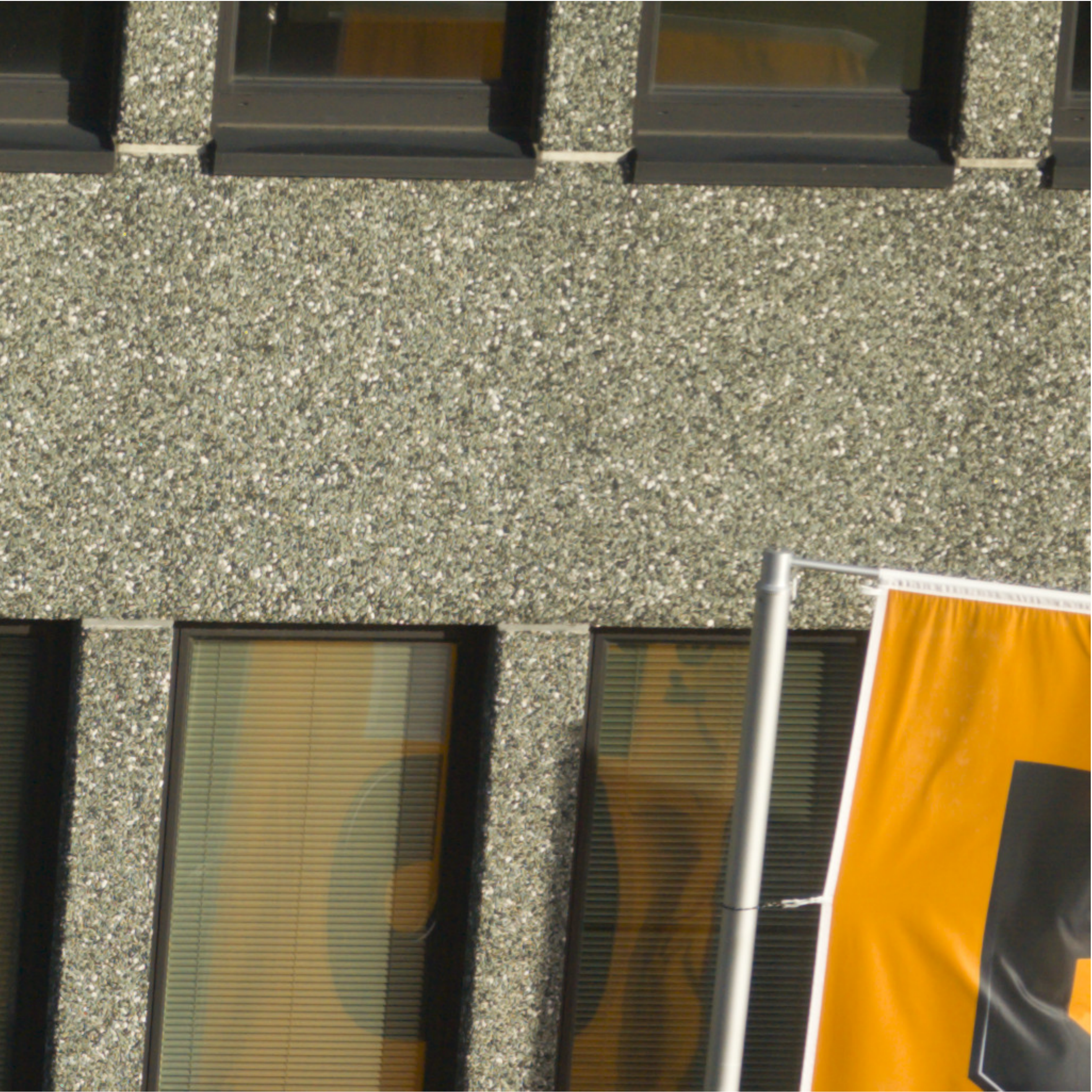} \hspace{-0.1cm}
\includegraphics[scale=0.14]{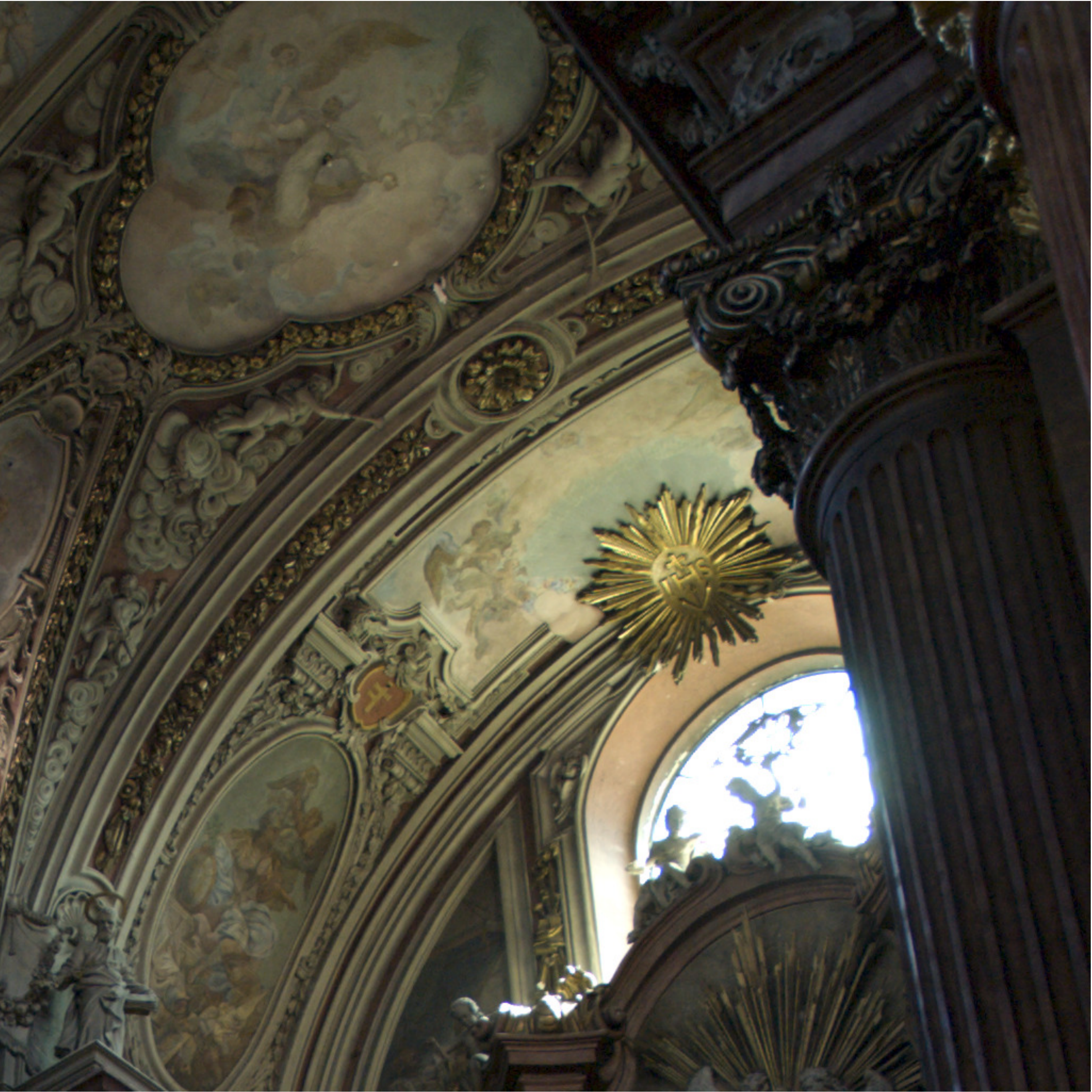}\hspace{-0.1cm}\\
\includegraphics[scale=0.14]{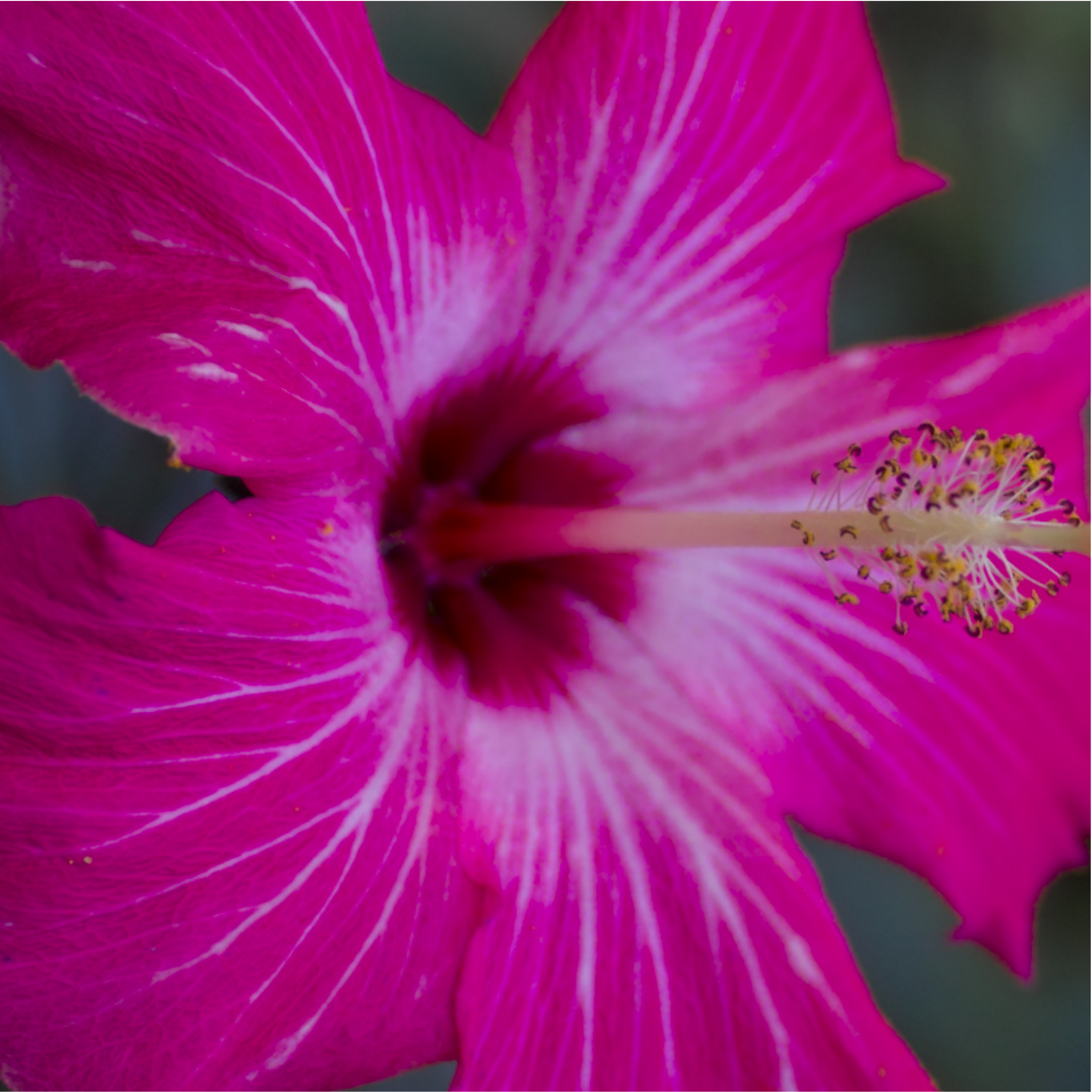} \hspace{-0.1cm}
%\hspace*{0.1cm}
\includegraphics[scale=0.14]{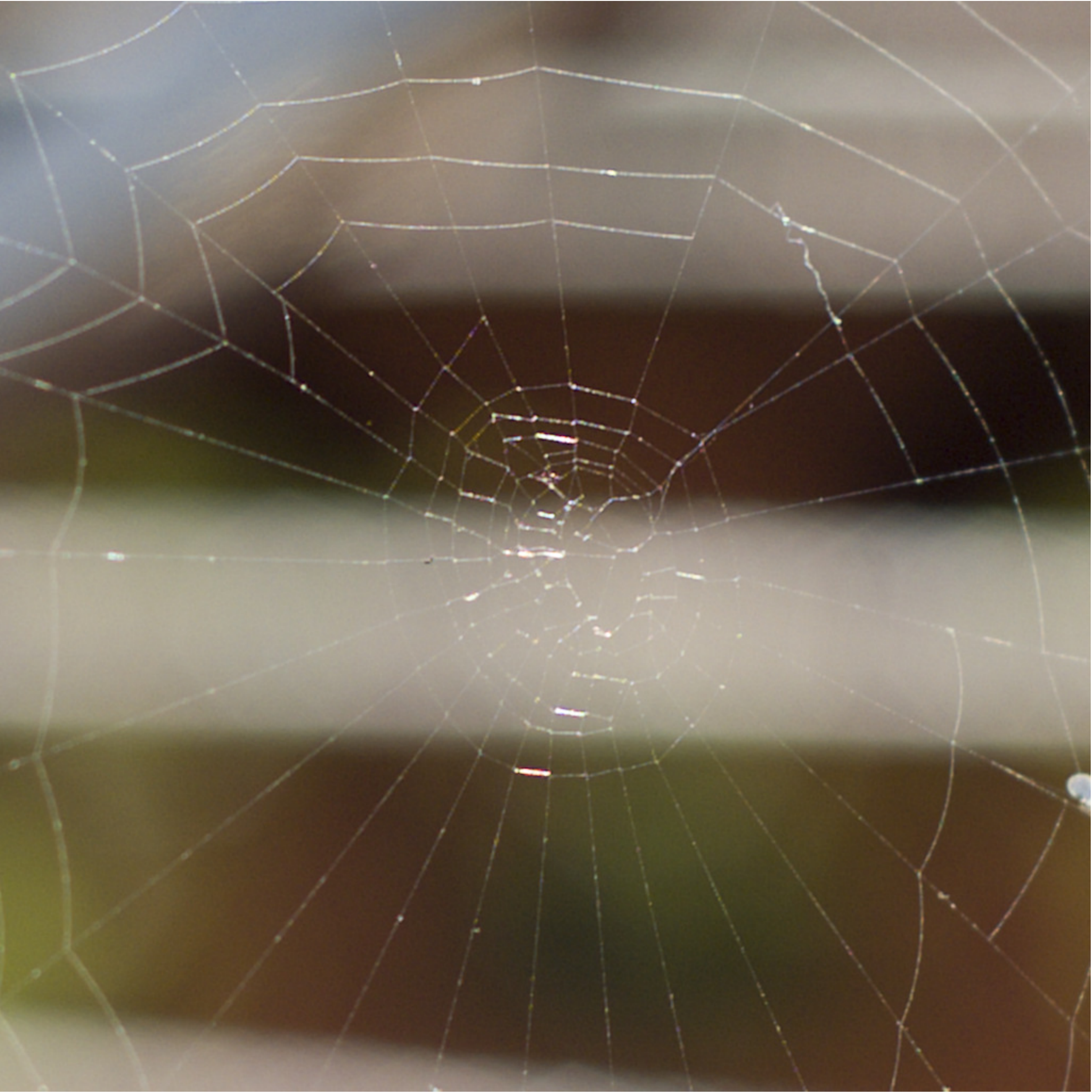} \hspace{-0.1cm}\\
\includegraphics[scale=0.14]{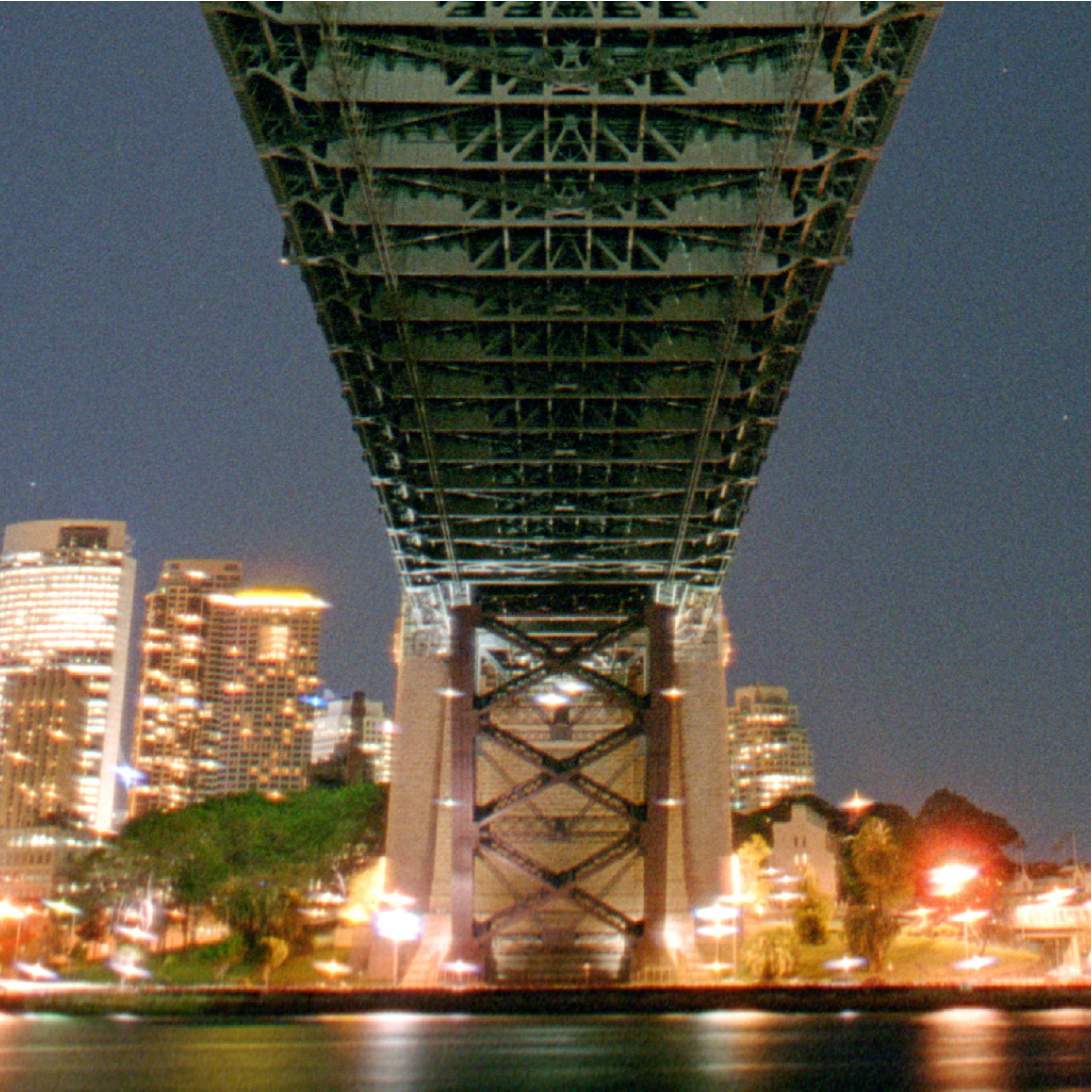} \hspace{-0.1cm}
\includegraphics[scale=0.14]{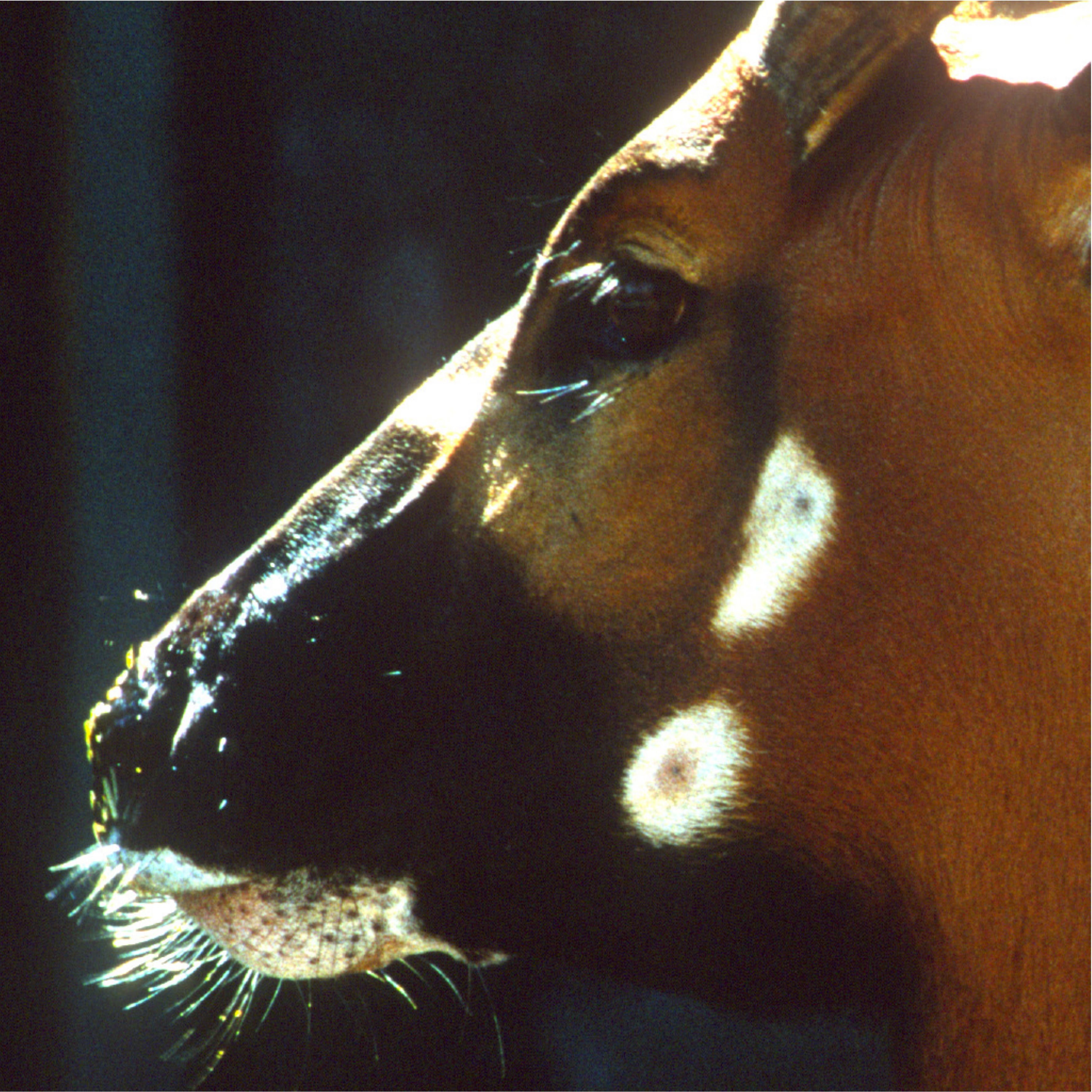} \hspace{-0.1cm}

\caption{Illustration of the $1024 \times 1024 \times 3$  panels from the six high resolution test images (No 10--15) listed in Table ~\ref{TABLE3}.}
\label{hsi}
\end{figure}

For realising the comparison we proceed as follows:
 we set the required 
value of PSNR as that produced by JPEG at quality=95
and tune compression with the other methods to produce 
the same PSNR. In our codec the tuning is realised by approximating the image up to $\PSNR_o=1.025 \cdot \PSNR$ (where 
$\PSNR$ is the targeted quality) and setting the 
quantisation parameter $\Delta$ so as to reproduce the 
targeted quality.
For compression with JPEG and JPEG2000 we use the 
MATLAB {\tt{imwrite}} command.  The comparison 
with WebP  was realised using the software for 
Ubuntu distributed on \cite{WebPpage}.

\begin{table}[h]
\begin{center}
\begin{tabular}{|r||l|r|c |}
\hline\hline
No& \quad Image & Size \,\quad\quad\quad &  time (secs) \\ \hline \hline
1 & Lenna &  $512 \times 512 \times 3$ & 1.8 \\ \hline
2 & Goldhill& $576 \times 720 \times 3$& 3.8 \\ \hline
3 & Barbara & $576 \times 720 \times 3$& 3.2 \\ \hline
4 & Baboon  & $512\times 512 \times 3$& 2.7 \\ \hline
5 & Zelda   & $576 \times 784 \times 3$& 2.5\\ \hline
6 & Sailboat & $512 \times 512 \times 3$& 1.8 \\ \hline
7 & Boy  &     $512 \times 768 \times 3$& 3.8 \\ \hline
8 & Jupiter &  $1072 \times 1376 \times 3$& 2.9 \\ \hline
9 & Saturn &  $1200 \times 1488 \times 3$&  2.6\\ \hline
10 &Building & $1024 \times 1024 \times 3$& 9.5\\ \hline
11 &Cathedral& $1024 \times 1024 \times 3$&7.6 \\ \hline
12 &Flower  &  $1024 \times 1024 \times 3$&3.6 \\ \hline
13 &Spider-web &$1024 \times 1024 \times 3$&3.8\\ \hline
14 &Bridge  &  $1024 \times 1024 \times 3$&8.5 \\ \hline
15 &Deer  &$1024 \times 1024 \times 3$&5.8\\ \hline  \hline
\end{tabular}
\end{center}
\caption{Test Images. The last column gives the approximation  times to produce the results in Table~\ref{TABLE4}}
\label{TABLE3}
\end{table}

\begin{table}[h]
\begin{center}
\begin{tabular}{||r|r|r|r|r|r||}
\hline \hline
I&dB&$\bjp$&$\bweb$&$\bjpp$&$\bsr$ \\\hline \hline
1& 35.9& 3.28& 2.21 &1.40 &\bf{1.37}\\ \hline
2& 36.6& 3.63& 2.47 &2.11 &\bf{2.01} \\ \hline
3& 37.2& 3.67& 2.80 &1.91 & \bf{ 1.77} \\ \hline
4& 28.8& 5.80& 4.66 &2.74 & \bf{2.48} \\ \hline
5& 39.3& 2.61& 1.81 &{\bf{1.08}} & \bf{1.07} \\ \hline
6& 30.9& 4.49& 3.21 &1.62 &\bf{ 1.51} \\ \hline
7& 32.6& 4.34& 3.07 &2.36 & \bf{2.22}  \\ \hline
8& 48.2& 0.60& 0.51 &0.24 & \bf{0.20}  \\ \hline
9&49.0& 0.34 &0.36&0.15 & \bf{0.12} \\ \hline
10& 37.4& 3.41& 2.35 & 1.84 &\bf{1.75} \\ \hline
11& 38.5& 2.84& 1.83 &  1.26 & \bf{1.20} \\ \hline
12& 41.5& 1.78& 1.08 & 0.56  & \bf{0.53}  \\ \hline
13& 34.5& 3.57& 2.48 & 1.61  &\bf{1.48} \\ \hline
14& 45.0& 1.55& 1.10 &\bf{0.58} &\bf{0.57} \\ \hline
15& 30.9& 3.76& 2.59 &1.04  &\bf{0.90} \\ \hline
 \hline
\end{tabular}
\end{center}
\caption{Compression rate ($\bpp$) corresponding to JPEG
($\bjp$), WebP ($\bweb$), JPEG2000 ($\bjpp$) and the proposed sparse representation ($\bsr$), for the
 values of PSNR given in the $2\mathrm{nd}$ column.}
\label{TABLE4}
\end{table}

\section{Conclusions}
\label{Conclu}
The application of a cross color transformation for
enhancing sparsity in the atomic decomposition of
RGB images has been proposed. It was demonstrated that
the effect of the transformation is to re-distribute the
most significant values in the dwt of the
2D channels. As a result, when approximating the
arrays by disregarding the less significant
entries, the quality of the reconstructed image
improves with respect to disabling the cross color
transformation.  Four transformations have been
considered: (i) a 3 point dct, (ii) 
the reversible YCbCr color space transform, 
(iii) the PC transform, 
(iv) a transformation learned from an independent
set of images. 

The quality of the image approximation was improved
further by approximating the transformed arrays by
an atomic decomposition using a separable dictionary 
and the  greedy pursuit
strategy  HBW-OMP2D.  The dct was singled out as the 
most convenient cross color transformation for approximating 
RGB color images in the wavelet domain. 

The approximation approach has been shown
to be relevant for image compression. By means of a
simple coding strategy the achieved compression considerably
improves upon the compression standards JPEG and WebP.
On a set of 15 typical test images the results are at least
as good as those produced by JPEG2000.
We feel confident that the findings communicated here will
motivate the consideration of dct as a cross color
transformations for other image processing applications which also benefit from the sparsity of a representation.

\end{document}